\documentclass[nosumlimits]{amsart}
\usepackage{amsmath}
\usepackage{amssymb}
\usepackage{layout}

\usepackage{graphicx}

\newtheorem{thm}{Theorem}[section]
\newtheorem{defin}[thm]{Definition}

\newtheorem{cor}[thm]{Corollary}

\numberwithin{equation}{section}

\newenvironment{remark}[1]{\medskip\par\noindent\small\
\begin{center}\textbf{Remark}\end{center}\begin{quote} #1}
{\medskip\par\noindent\end{quote}}

\newcommand{\B}{\mathcal{B}}

 \newcommand{\less}{\setminus}
 
 \newcommand{\set}[1]{\left\{#1\right\}}

 \newcommand{\norm}[1]{\left\Vert#1\right\Vert}
 
 \newcommand{\qtext}[1]{\quad\text{#1}\quad}
 \newcommand{\fa}{\qtext{for all}}
 \newcommand{\bb}{\begin{equation*}}
 \newcommand{\ee}{\end{equation*}}
 \newcommand{\bp}{\begin{proof}}
 \newcommand{\ep}{\end{proof}}

\begin{document}

\title[Fields generated by plasma flows]{Electromagnetic
field generated\\ by plasma flows\\
and Feynman and Li\'{e}nard-Wiechert formulas\\ for
a moving point charge }
\bigskip

\author{ Victor M. Bogdan }

\address{Department of Mathematics, McMahon Hall 207, CUA, Washington DC 20064, USA}

\date{7 July 2009}

\email{bogdan@cua.edu}


\begin{abstract}
This note represents a stepping stone from the discovery of the precise mathematical
formula for electromagnetic field generated by a moving point charge, the amended
Feynman formula, see Bogdan \\
http://arxiv.org/abs/0909.5240, and
leading to the to the general formula of gravitational and
electromagnetic fields generated by moving matter in
a Lorentzian frame of special theory of relativity, see\\
http://arxiv.org/abs/0910.0538.

    In this note the author introduces the notion
of flow of matter in a Lorentzian frame.
This notion is relativistic in the sense of Einstein's special theory of relativity.

    The author presents explicit formulas
suitable for a digital computer permitting one to find
time delay for an action from a flow line to any
point in the Lorentzian frame.
The time delay field for any flow of plasma in a fixed Lorentzian
frame is unique.

    Using this field he introduces the retarded time field
and fundamental fields corresponding to the flow with
a free parameter representing the initial positions of the
lines of flow.

    By means of these fields one can represent and establish relations
between wave, Lorentz gauge, and Maxwell equations,
and Lienard-Wiechert potentials, and amended Feynman's formula.

    The initial distribution of charges over the initial position
of the flow is given by a signed measure of finite variation
defined over Borel sets. It may include discrete and continuous components.
\end{abstract}


\maketitle

This note represents a stepping stone from the discovery of the precise mathematical
formula for electromagnetic field generated by a moving point charge, the amended
Feynman formula, see Bogdan \cite{bogdan65}, and
leading to the to the general formula of gravitational and
electromagnetic fields generated by moving matter in
a Lorentzian frame of special theory of relativity, Bogdan \cite{bogdan71}.

The the generalized densities of mass and of charge in the above fields may consist
of discrete and continuous components not like in general theory
of relativity, for reference see Dirac \cite{dirac}.

As a byproduct of the paper \cite{bogdan71}, for the case when the charge field
has a Lebesgue summable density,
one obtains precise electromagnetic potentials that correct Feynman's formulas
presented in Feynman-Leighton-Sands \cite{feyn}, vol. 2, chapter 15, page 15.

The results contained in this note have been presented at the annual meeting of
    the Society for Applied and Industrial Mathematics held in Denver,
    Colorado, on 7 July 2009 \cite{bogdan71}.

Let us get to the essentials. For the sake of simplicity in notation we
select units so that the speed of light is $c=1$ and the electrostatic
constant satisfies the condition $4\pi\epsilon_0=1.$

We are working here
in the environment of Einstein's special theory of relativity
in a fixed Lorentzian frame.

\bigskip
\section{Uniqueness of time delay field for a flow of plasma}
\bigskip

\bigskip

\begin{defin}[Plasma flow]
Let $F\subset R^3$ be a compact set representing the position of plasma
at some initial time $t_0.$

By a plasma flow we shall understand a continuous function $r_2(t,r_0)$ from the
product $R\times F$ into $R^3,$ having third derivative with respect to $t$ and
such that the velocity $v_2(t,r_0)=\dot{r}_2(t,r_0)$ and the acceleration
$a_2(t,r_0)=\dot{v}_2(t,r_0)$ and the derivative $\dot{a}_2(t,r_0)$
are continuous on the product  $R\times F.$

Moreover the following two conditions are satisfied:
\begin{enumerate}
    \item For every time $t_1$ there is a velocity
$v_1<c=1$ such that
\begin{equation}\label{velocity v 1}
    |v_2(t,r_0)|\le v_1\fa t\le t_1\text{ and } r_0\in F.
\end{equation}

    \item  For every time $t\in R$ the map $P_t$ given by the formula
\begin{equation}\label{P t map}
    P_t(r_0)=r_2(t,r_0)\fa t\in R\text{ and } r_0\in F
\end{equation}
represents a homeomorphism of $F$ onto $P_t(F).$
\end{enumerate}
\end{defin}

\bigskip

Clearly we have
\begin{equation*}
    P_{t_0}(r_0)=r_2(t_0,r_0)=r_0\fa r_0\in F.
\end{equation*}
The function $t\mapsto r_2(t,r_0)$ will be called a line of flow corresponding to
the index $r_0.$

\bigskip

Let $T=T(r_1,t,r_0)$ denote the time delay required to reach
point $r_1\in R^3$ at time $t$ from the line of flow corresponding to
index $r_0.$ Its value must satisfy the Lorentz time delay equation
\begin{equation}\label{time delay}
    T=|r_1-r_2(t-T,r_0)|.
\end{equation}

\bigskip

\begin{thm}[Time delay is unique and continuous]
For every point $r_1\in R^3,$ and time $t\in R,$ and index $r_0\in F$
there exists one and only one solution $T$ of equation (\ref{time delay}).
Moreover the function $T=T(r_1,t,r_0)$ is continuous on its entire domain
$R^3\times R\times F.$
\end{thm}

\bigskip
Notice the following relations
\bb
    T=0\Leftrightarrow\{ r_1=r_2(t,r_0)\text{ for some }r_0\in F\}
    \Leftrightarrow r_1\in P_t(F).
\ee
The set $P_t(F)$ represents the position of the plasma at time $t.$

Now define sets
$$
    G=\set{(r_1,t,r_0)\in R^3\times R\times F:\ T(r_1,t,r_0)>0}
$$
and
\bb
    G_0=\set{(r_1,t)\in R^3\times R:\ T(r_1,t,r_0)>0\fa r_0\in F}.
\ee

\bigskip

\begin{thm}[Sets $G$ and $G_0$ are open]
The set $G$ is nonempty and open in the product space $R^3\times R\times F$
and so is the set $G_0$
in the product space $R^3\times R.$
\end{thm}

\bigskip

\bigskip
\section{Fundamental fields corresponding to the flow}
\bigskip

\begin{defin}[Fundamental fields]
Introduce the retarded time function
\begin{equation*}
 \tau=\tau(r_1,t,r_0)=t-T(r_1,t,r_0)\fa (r_1,t,r_0)\in R^3\times R\times R^3,
\end{equation*}
retarded velocity
\begin{equation*}
 v=v_2(\tau(r_1,t,r_0),r_0)\fa (r_1,t,r_0)\in R^3\times R\times R^3,
\end{equation*}
 and retarded acceleration
\begin{equation*}
 a=a_2(\tau(r_1,t,r_0),r_0)\fa (r_1,t,r_0)\in R^3\times R\times R^3,
\end{equation*}
and vector field $r_{12}$ by
\begin{equation*}
 r_{12}=r_1-r_2(\tau(r_1,t,r_0),r_0)\fa (r_1,t,r_0)\in R^3\times R\times R^3.
\end{equation*}
Introduce the unit vector field $e,$ and the fields $u$ and $z$
by the formulas
\begin{equation}\label{unit vector field}
    \qtext{and}e=\frac{r_{12}}{T}\qtext{and} u=\frac{1}{T}
    \qtext{and}z=\frac{1}{(1-\langle e,v\rangle)}\qtext{on} G.
\end{equation}
These functions will be called the {\bf fundamental fields} associated
with the flow $r_2(t,r_0),$ where $t\in R$ and $r_0\in F.$
\end{defin}

Notice that by the definition of flow of plasma
the velocities are smaller in magnitude than
the speed of light $c=1.$ Thus we must have for the dot product
$|\langle e,v\rangle|\le|v|<1.$
So the field $z$ is well defined.

All the above functions consist of compositions of continuous functions,
therefore each of them is continuous on its respective  domain and thus
all of them are continuous on their common domain, the set $G.$

\bigskip

We would like to stress here that the fundamental fields depend on the Lorentzian
frame, in which we consider the trajectory.
It is important to find expressions involving fundamental fields that
yield fields invariant under Lorentzian transformations.

Lorentz \cite{lorentz} and Einstein \cite{einstein2a}, Part II, section 6, established that
fields satisfying Maxwell equations are invariant under Lorentzian transformations.
\bigskip

Our main goal is to prove that fields constructed for flows of plasma
will satisfy Maxwell equations. We shall do this by showing that these fields
are representable by means of fundamental fields and using the formulas
for partial derivatives of the fundamental fields prove that
such fields generate fields satisfying Maxwell equations.

Introduce operators $D=\frac{\partial}{\partial t}$ and
$D_i=\frac{\partial}{\partial x_{i}}$ for $i=1,2,3$ and
$\nabla=(D_1,D_2,D_3).$

Observe that $\delta_i$ in the following formulas denotes the i-th
unit vector of the standard base in $R^3$ that is
$\delta_1=(1,0,0),$ $\delta_2=(0,1,0),$ $\delta_3=(0,0,1).$

The symbols $e_i,$ $v_i,$ $a_i,$ denote the corresponding component
of the vector fields $e,$ $v,$ $a,$ respectively.

\bigskip

\begin{thm}[Partial derivatives of fundamental fields]
Assume that in some Lorentzian frame
we are given a plasma flow $(t,r_0)\mapsto r_2(t,r_0).$
For partial derivatives with respect to coordinates of the vector $r_1$
we have the following identities on the set $G$
\begin{eqnarray}
\label{DiT}       D_iT&=&ze_i  ,\\
\label{Diu}       D_iu&=&-zu^2e_i  ,\\
\label{Div}          D_iv&=&-e_iza ,\\
\label{Di tau}    D_i\tau&=&-ze_i,\\
\label{Die}         D_ie&=& -uze_ie+u\delta_i+uze_iv \qtext{where} \delta_i
                    =(\delta_{ij}),\\
\label{Diz}          D_i z &=& -z^3e_i \langle e,a \rangle
                -uz^3e_i+ uz^2e_i+uz^2v_i+uz^3e_i \langle v,v \rangle \\
\label{grad T}\nabla T&=&ze  ,\\
\label{grad u}\nabla u&=&-zu^2e ,\\
\label{grad z} \nabla z&=&  -z^3 \langle e,a \rangle e-uz^3e+ uz^2e+uz^2v
                +uz^3 \langle v,v \rangle e.
\end{eqnarray}
and for the partial derivative with respect to time we have
\begin{eqnarray}
\label{DT}    DT&=&1-z,\\
\label{Du}    Du&=&zu^2-u^2,\\
\label{D tau}    D\tau&=&z,\\
\label{Dv}    Dv&=&za  ,\\
\label{De}    De&=&-u e+ u z e-u z v,\\
\label{Dz}    Dz&=&uz-2uz^2 +z^3 \langle e,a \rangle +uz^3-uz^3 \langle v,v \rangle .
\end{eqnarray}
Since the expression on the right side of each formula represents
a continuous function, the fundamental fields are at least of class $C^\infty$ on the set $G.$
\end{thm}
%
\bigskip
The proof of the above theorem  is similar to the proof of analogous theorem in
Bogdan \cite{bogdan65}.

\section{Integration with respect to a signed measure}

Let $V$ be a prering of subsets of $F$
consisting of sets of the form $Q\cap B$ where $Q$ is compact and $B$
is open. See Bogdanowicz \cite[page 498]{bogdan10} available on the web.

Assume that the set functions $q^+(A)$ and $q^-(A)$ represent, respectively,
the total positive and total negative charge  contained in the body covered by
the set $A\in V.$
We shall assume that these functions are countably
additive.

\begin{remark}
A heuristic argument relying on assumption
that charge of an electron is indivisible can be presented as follows:
Take a decomposition of a set $A\in V$
into a countable union of disjoint sets
\bb
    A=A_1\cup A_2\cup\ldots A_n\cup\ldots.
\ee
Since every charge comes in the form of finite number of indivisible unit charges,
that are all equal to the charge of a single electron,
only a finite number of the sets may contain
a charge. Thus starting from a sufficiently large index $n_0$ all sets $A_n$
will have charge zero. Thus
\bb
    q^+(A)=\sum_{n\le n_0}q^+(A_n)+\sum_{n> n_0}0=\sum_{n=1}^\infty q^+(A_n).
\ee
Similarly we can get countable additivity of $q^-.$
\end{remark}

Put
$q(A)=q^+(A)+q^-(A)$ and $\eta(A)=q^+(A)-q^-(A).$
The value $q(A)$ represents the total charge in the body covered by the
set $A$ and $\eta(A)$
represents a non-negative countably additive set function on $V$
such that $|q(A)|\le \eta(A).$

Such a function satisfies the requirements of a volume function
as defined in \cite[page 492]{bogdan10}. Observe that the
function $q$ belongs to the space $M,$ defined on page 492, and its norm $\norm{q}\le 1.$
Therefore we can use the trilinear integral $\int u(f,dq)$ developed there. In our case
for the bilinear operator $u(y,r)=ry=yr$ defined for $y\in Y$ and $r\in R,$
where $Y$ stands for either the vector space $R^3$ or the space of $R$ of reals.

Thus we can use the theory
developed in the papers Bogdanowicz \cite{bogdan10} and \cite{bogdan14}. The latter one is
also available on the web.
These two papers provide all the tools of Lebesgue and Bochner
theory on measure and integration, based on measure on sigma rings of sets,
available if needed in applications.

\bigskip
Concerning notation: We are using the symbol $\int u( f,dq)$ to denote the integral
over the entire space $F$ of integration. When it is desirable to indicate the variable
of integration we shall write $\int u(f(r_0),q(dr_0)).$

If we have a set $A\subset F$
and a function $f:F\mapsto Y$ such that the product $\chi_Af$ yields an
$\eta$-summable function,
where $\chi_A$ denotes the characteristic function of the set $A,$
then we shall say that the function $f$ is summable on the set $A$ and by its
integral over the set we shall understand the following
\begin{equation*}
    \int_A u(f,dq)=\int u(\chi_Af,dq).
\end{equation*}
Since $\chi_Ff=f$ for all functions defined on $F,$
the two notions for the set $F$ coincide,  that is
\begin{equation*}
    \int u(f,dq)=\int_F u(f,dq).
\end{equation*}

In the case when the bilinear form $u(r,\lambda)=r\lambda=\lambda r$ we shall
write the integral with respect to $u$ just as $\int f\,dq.$

\bigskip
From \cite{bogdan10}, Theorem 8, page 498, and Theorem 5, page 497,
we can get the following theorem.

\begin{thm}[Commutativity of differential and integral operators]
\label{Commutativity of differential and integral operators}
Assume that $h$ is either a scalar or a vector function on the
open set $G.$ If $h$ is continuous on $G$ and for every fixed
$r_0\in F$ the function $(r_1,t) \mapsto h(r_1,t,r_0)$ has partial
derivatives with respect to the coordinates of the point $(r_1,t)$
and these derivatives are continuous on the the set $G,$ then the function
\bb
    H(r_1,t)=\int_F h(r_1,t,r_0)\,q(dr_0)\fa (r_1,t)\in G_0
\ee
is well defined and has continuous  partial derivatives on $G_0.$

Moreover we have the following formulas
\bb
    D\int_F h\,dq=\int_F Dh\,dq\qtext{and}D_i\int_F h\,dq=\int_F D_ih\,dq.
\ee
\end{thm}
\bigskip

\section{Fields with free parameter representing index of the line of flow}
\bigskip

In the following $r_0,$ index of the line of flow,
represents a free parameter from the compact
set $F.$ All the  partial derivatives are with respect to coordinates
of the point $(r_1,t)\in R^3\times R.$

\begin{thm}\label{wave equations}
On the set  $G=\set{(r_1,t,r_0):\ T(r_1,t,r_0)>0}$ we have the following identities
involving wave equations and Lorentz gauge equation
\begin{equation*}
    (\nabla^2-D^2)[uz]=0,\quad(\nabla^2-D^2)[uzv]=0,\quad
    \nabla\cdot[uzv]+D[uz]=0
\end{equation*}
\end{thm}

\bigskip

\begin{thm}[Wave equation with gauge imply Maxwell equations]
\label{wave-->Maxwell}
Assume that on the set
\begin{equation*}
 G=\set{(r_1,t,r_0):\ T(r_1,t,r_0)>0}
\end{equation*}
there are given two scalar fields $\phi$ and $S$ and two vector fields $A$ and $J.$

Assume that the fields $\phi$ and $A$ have second partial derivatives with respect to the
coordinates of the point $(r_1,t)$ and these derivatives are continuous on the set
$G.$

If these fields satisfy the following wave equations with Lorentz gauge formula
\begin{equation}
    \nabla^2\phi-\frac{\partial^2}{\partial t^2}\phi=-S,
    \quad\nabla^2A-\frac{\partial^2}{\partial t^2}A=-J,
    \quad \nabla\cdot A+\frac{\partial}{\partial t}\phi=0\qtext{ on the set }G,
\end{equation}
then the fields $E$ and $B$ defined by the formulas
\begin{equation*}
    E=-\nabla \phi - \frac{\partial}{\partial t} A\qtext{and}B
    =\nabla\times A\qtext{ on the set }G,\
\end{equation*}
will satisfy the following Maxwell equations
\begin{equation}\label{Maxwell equations for potentials}
    (a)\quad\nabla\cdot E=S,\ \quad(b)\quad\nabla\times E
    =- \frac{\partial}{\partial t} B,\ \quad(c)\quad
    \nabla\cdot B=0,\ \quad(d)\quad\nabla\times B
    =\frac{\partial}{\partial t}E+J
\end{equation}
on the set $G.$
\end{thm}

\bigskip

\begin{thm}On the set  $G=\set{(r_1,t,r_0):\ T(r_1,t,r_0)>0}$
introduce fields $\phi$ and $A$ by  Li\'{e}nard-Wiechert formulas
\begin{equation*}
 \phi=uz\qtext{and}A=uzv
\end{equation*}
and define fields
 $E,$ and $B,$ by the formulas
\begin{equation*}
 E=-\nabla \phi-DA\qtext{and}B=\nabla\times A.
\end{equation*}
and the fields $E_f$ and $B_f$ by the  formulas
\begin{equation*}
 E_f=u^2e+u^{-1}D(u^2e) +D^2e\qtext{and}B_f=e\times E.
\end{equation*}
Then the following Maxwell equations are satisfied on the entire set $G$
\begin{equation}\label{Maxwell equations-ex}
    \nabla\cdot E=0,\quad\nabla\times E=-DB,\quad
    \nabla\cdot B=0,\quad\nabla\times B=DE
\end{equation}
and for the fields $E$ and $B,$ we have the following representations
\begin{equation*}
    \begin{split}
    E_f&=E=u^2e+u^{-1}D(u^2e)+D^2(e) \\
     &=-uz^2a  +uz^3\langle  e,a\rangle  e-uz^3\langle  e,a\rangle  v
     +u^2z^3e-u^2z^3\langle  v,v\rangle  e-u^2z^3v+u^2z^3\langle  v,v\rangle  v,\\
    B_f&=B=-uz^2\ e\times a -uz^3 \langle e,a \rangle \ e \times v-u^2z^3\ e \times v
        +u^2z^3 \langle v,v \rangle \ e \times v\\
    \end{split}
\end{equation*}
at every point $(r_1,t,r_0)$ the set $G.$
\end{thm}

We shall remind the reader the meaning of the notation.
The set $P_t(F)$ represents the position of the plasma at time $t.$

We have defined sets
$$
    G=\set{(r_1,t,r_0)\in R^3\times R\times F:\ T(r_1,t,r_0)>0}
$$
and
\bb
    G_0=\set{(r_1,t)\in R^3\times R:\ T(r_1,t,r_0)>0\fa r_0\in F}.
\ee

Introduce here the set $\B$ by the formula
\begin{equation*}
     \B=\set{(r_1,t)\in R^3\times R:\ r_1=r_2(t,r_0)\text{ for some }r_0\in F}.
\end{equation*}

The set $\B$ will be called the {\bf trace of the flow}.
It follows from continuity of the function $r_2$ that the trace $\B$
represents a closed set.

Notice the relation
\begin{equation*}
     G_0=R^4\less \B.
\end{equation*}
This means that $G_0$ as a complement of the closed set $\B$ is open.

\bigskip

\section{Fields over the complement of the trace}
\bigskip

\begin{defin}[Scalar and vector potentials]
For any flow of plasma and any measure $q(Q)$ over $F$
define the {\bf scalar potential} $\phi,$ and
the {\bf vector potential} $A,$
are well defined on the open set $G_0$ by the formulas
\begin{equation}
    \begin{split}
        \phi(r_1,t)&=\int_F [(uz)(r_1,t,r_0)]\,q(dr_0),\\
        A(r_1,t)&=\int_F [(uzv)(r_1,t,r_0)]\,q(dr_0),\\
    \end{split}
\end{equation}
\end{defin}
\bigskip

Since for any point $(r_1,t)$ not in the trace of the flow the
functions under the integral sign are continuous with respect to $r_0,$
we can conclude
that the following is true.

\begin{thm}[Potentials are well defined]
For any flow of plasma and any measure $q(Q)$ over $F$
the scalar and vector potentials are well defined
on the set $G_0$ and represent continuous functions.
\end{thm}
\bigskip

\begin{thm}[Potentials and Maxwell's equations]
For any flow of plasma and any measure $q(Q)$ of finite variation over $F$
define fields $S$ and $J$ 
by the following  formulas
\begin{equation}
    \nabla^2\phi-\frac{\partial^2}{\partial t^2}\phi=-S,
    \quad\nabla^2A-\frac{\partial^2}{\partial t^2}A=-J.
\end{equation}

Then the fields defined by
$E=-\nabla\phi-\frac{\partial}{\partial t}A$ and $B=\nabla\times A$
will satisfy the Maxwell equations
\begin{equation}
    \nabla\cdot E=S,\ \quad\nabla\times E
    =-\frac{\partial}{\partial t}B,\ \quad
    \nabla\cdot B=0,\ \quad\nabla\times B
    =\frac{\partial}{\partial t}E+J
\end{equation}
and can be represented by means of the integral formulas
\begin{equation}\label{feynman's formula}
    E=\int_F \left(u^2e+u^{-1}\frac{\partial}{\partial t}(u^2e)
    +\frac{\partial^2}{\partial t^2}e\right)\,dq,
\end{equation}
\begin{equation*}
     B=\int_F e\times \left(u^2e+u^{-1}\frac{\partial}{\partial t}(u^2e)
    +\frac{\partial^2}{\partial t^2}e\right)\,dq,
\end{equation*}

Moreover the field $S$ represents the generalized density of charges and
the field $J$ represents the generalized density of currents.
They satisfy the equation of continuity
\begin{equation*}
 \nabla\cdot J+\frac{\partial}{\partial t}S=0
\end{equation*}
of flow of charge.
Here $F$ represents the initial position of the plasma in $R^3$,
and the scalar field $u$ and
the vector field $e$ are defined in formula (\ref{unit vector field}).
\end{thm}
\bigskip

\section{The independence of the fields from initial measure}
\bigskip
Assume that $q(A)$ represents as before the total charge contained in the body covered by a
set $A\subset F$ at time $t_0.$ Assume that at some later time $\tilde{t}_0$
the position of the plasma is in the set $\tilde{F}$ and
\begin{equation*}
 P(r)=r_2(\tilde{t}_0,r)\fa r\in F
\end{equation*}
represents transformation of points in $F$ at time $t_0$ to points in $\tilde{F}$
at time $\tilde{t}_0.$ Since by definition of a plasma flow the
transformation $P$ is homeomorphism the set $\tilde{F}$ is compact since $F$ is.

Let $\tilde{V}$ be the prering consisting of intersections of compact
sets with open sets of the space $\tilde{F}.$ Sets of this prering
can be represented as set differences of two compact sets.
Define set function
\begin{equation*}
 \tilde{q}(\tilde{A})=q(P^{-1}(\tilde{A}))\fa \tilde{A}\in \tilde{F}.
\end{equation*}
Since the transformation $P^{-1}$ preserves compact sets and set differences,
the set function $\tilde{q}$ is well defined. We shall prove that
it represents distribution of charges at time $\tilde{t}_0.$

The following theorem shows that the formulas for potentials in
integral form do not depend on transition from one initial time $t_0$
to another $\tilde{t}_0.$

\bigskip

\begin{thm}
Let $h(r)$ be a continuous function on the set $F$ with values in either the vector
space $R^3$ or the space $R$ of reals.
Let
\begin{equation*}
    \tilde{h}(r)=h(P^{-1}(r))\fa r\in \tilde{F}
\end{equation*}
Then we have the equality
\begin{equation}\label{initial time transition}
 \int_{\tilde{F}} \tilde{h} \,d\tilde{q}=\int_F h \,dq.
\end{equation}
\end{thm}

\bigskip

\begin{cor}
The scalar potential $\phi$ and the vector potential $A$ are independent
of the initial time $t_0$ when the distribution $q$ of charges was observed,
and as a consequence the electric field $E$ and the magnetic field $B$ also
do not depend on the initial time when the distribution was observed.
\end{cor}

\bigskip



\begin{thebibliography}{99}

\bibitem{bogdan10} W.M.~Bogdanowicz, (also known as V.M.~Bogdan),
    {\em A Generalization of the
    Lebesgue-Bochner-Stieltjes Integral and a New Approach to the
    Theory of Integration}, Proc. of Nat. Acad. Sci. USA,
    Vol.~53, No.~3, (1965), p.~492--498\\
    (On the web: http://faculty.cua.edu/bogdan/rep/10.pdf).
\bibitem{bogdan14}----------- 
    {\em An Approach to the Theory of
    Lebesgue-Bochner Measurable Functions and to the Theory of
    Measure}, Math. Annalen 164, (1966), p.~251--269.\\
    (On the web: http://faculty.cua.edu/bogdan/rep/14.pdf).

\bibitem{bogdan61}V.M.~Bogdan,
    {\em Existence of Solutions to Differential Equations of
    Relativistic Mechanics Involving Lorentzian Time Delays,}
    Journal of Mathematical Analysis and Applications,
    Academic Press, 118, No. 2, September (1986), p. 561-573.\\
    (On the web: http://faculty.cua.edu/bogdan/rep/61.pdf).
%

\bibitem{bogdan64}V.M.~Bogdan, {\em Feynman's Electromagnetic
    Fields Induced by Moving Charges
    and the Existence and Uniqueness of Solutions
    to N-Body Problem of Electrodynamics,}
    Quaestiones Mathematicae, vol. 32, (2009):1-87

%

\bibitem{bogdan65}V.M.~Bogdan, {\em Fields generated by a moving
    relativistic point mass and mathematical correction to Feynman's law}
    (to appear in Quaestiones Mathematicae),
    (On the web http://arxiv.org/abs/0909.5240)

%
\bibitem{bogdan71}V.M.~Bogdan,
    {\em Electromagnetic Field Generated by a Moving Plasma
    and Feynman and Li\'{e}nard-Wiechert Formulas for
    a Moving Point Charge,} presented at annual meeting of
    the Society for Applied and Industrial Mathematics, Denver,
    Colorado, 7 July 2009

\bibitem{bogdan72}V.M.~Bogdan,
    {\em Relativistic gravity fields and electromagnetic fields
    generated by flows of matter}, (these archives)
    http://arxiv.org/abs/0910.0538


\bibitem{dirac} P.A.M. Dirac, {\em General Theory of Relativity}
    Princeton University Press, Princeton, N.J., 1996




\bibitem{einstein2a} A. Einstein, {\em  Zur Elektrodynamik bewegter K\={o}rper,} Annalen der
    Physik. 17, 891: 1905.\\
    Enlish translation: {\em On the Electrodynamics of Moving Bodies,}
    The Principle of Relativity, Methuen and Company, 1923
    and on the web at\\
    http://www.fourmilab.ch/etexts/einstein/specrel/www/



\bibitem{feyn} R. P. Feynman, R. B. Leighton, and M. Sands,
    {\em The Feynman Lectures on Physics,}
    Vol. 1--3, Addison-Wesley, Reading, Mass., 1975.



\bibitem{lorentz} H.A. Lorentz,
    {\em Lectures on theoretical physics,}
    MacMillan \& Co., London, 1927

\end{thebibliography}
\end{document}